\documentclass[10pt]{article}
\usepackage{CJK}
\usepackage{amsmath}
\usepackage{amsfonts}
\usepackage[square,numbers,sort&compress]{natbib}
\usepackage{amsmath,amssymb,array,graphicx}
\usepackage{xcolor}
\usepackage{ifpdf}
\usepackage{epstopdf}
\usepackage{epsfig}
\usepackage{float}
\usepackage{booktabs}
\usepackage{hyperref}

\usepackage{subfigure}
\usepackage{algorithm}
\usepackage{algorithmic}
\usepackage{multirow}
\usepackage{float}

\newtheorem{definition}{Definition}[section]
\newtheorem{property}{Property}[section]

\newtheorem{remark}{Remark}

\newtheorem{example}{Example}

\newcommand{\beq}{\begin{equation}}
\newcommand{\eeq}{\end{equation}}
\newcommand{\bey}{\begin{eqnarray}}
\newcommand{\eey}{\end{eqnarray}}

\def\ds{\displaystyle}

\def\min{{\mbox{min}}}
\usepackage{setspace}

\begin{document}

\newpage
\title{The Reconstruction and Prediction Algorithm of the Fractional
TDD for the Local Outbreak of COVID-19}
\author{
Yu Chen \thanks{School of Mathematics, Shanghai University of Finance and Economics, Shanghai, China.(\tt chenyu@usfc.edu.cn)}
\and
Jin Cheng \thanks{School of Mathematical Sciences, Fudan University, Shanghai, China. (\tt jcheng@fudan.edu.cn)}
\and
Xiaoying Jiang
\thanks{School of Mathematical Sciences, Zhejiang University, Hangzhou 310027, China.({\tt jiangxiaoying@zju.edu.cn})}
 \and
Xiang Xu
\thanks{Corresponding author. School of Mathematical Sciences, Zhejiang University, Hangzhou 310027, China.({\tt xxu@zju.edu.cn})
}
}

\date{}
\maketitle

\begin{abstract}
From late December, 2019, the novel Corona-Virus began to spread in the mainland of China. For predicting the trend of the Corona Virus spread, several time delay dynamic systems (TDD) are proposed. In this paper, we establish a novel fractional time delay dynamic system (FTDD) to describe the local outbreak of COVID-19. The fractional derivative is introduced to account for the sub-diffusion process of the confirmed and cured people’s growth. Based on the public health data by the government, we propose a stable reconstruction algorithm of the coefficients. The reconstructed coefficients are used to predict the trend of the Corona-Virus. The numerical results are in good agreement with the public data.
\end{abstract}


\smallskip\noindent
{\bf Keywords.} COVID-19, fractioal derivative,  time delay dynamic systems
\section{Introduction}

In late December 2019, the novel coronavirus COVID-19 emerged in Wuhan, the captial of Hubei province and one of largest cities in the central part of China. One distinct feature of the COVID-19 virus is that it can spread from person to person as confirmed in \cite{Chan}. There were more than 74,000 confirmed infections and more than two thousands people died until February 20, 2020 \cite{wz}. Moreover, the situations in Japan and South Korea are becoming serious as well. Hence, predicting the further spread of the COVID-19 epidemic attracts a major public attention.

Considering the epidemic's feature of spreading during the latent period, Chen \cite{Chen,Yan} et al. applied the time delay process to describe the typical featrue and proposed a novel dynamical system to predicte the outbreak of COVID-19. Based on the daynamical system, Chen et al. \cite{Chen1} also proposed a time delay dynamic system with external source to describe the trend of local outbreak for the COVID-19 and provided numerical simulations. Mathematically speaking, in their models, the time delay term involving an integral with smooth kernel could be viewed as a special kind of fractional integral in some sense.

The fractional derivatives \cite{Podlubny} provide useful tools for the description of memory and hereditary properties of different materials. And the differential equations with fractional derivatives stand out in studying fractal geometry and fractal dynamics \cite{Carpinterj,Mandelbrot}. As the fractional derivative can describe the anomalous diffusion, in this paper, we propose a novel time delay dynamic system with fractional order, which is based on the model in \cite{Chen}. In the newly proposed system, the Riemann-Liouville derivative is added which can describe the confirmed and cured people's growth process.

The rest of paper is organized as follows. Section 2 presents notaions, assumptions and the corresponding fractional time delay dynamic system. The approach for estimating the future number of diagnosed people with this system is provided in section 3. Based on the offical data given by CCDC everyday, serveral numerical examples are exhibted in section 4 to verify the rationality of the fractional model and effectiveness of the estimation scheme, and to give the resonable range of the fractional order. Finally, we present some concluding remarks in section 5.

\section{Formulation of Fractional Time Delay Dynamic System}
Before proceeding to a fractional time delay dynamic system, we first introduce some definitions, notaions and assumptions.
\begin{definition}
	The left side Riemann-Liouville derivative is defined as \cite{Podlubny}
\begin{equation}\label{01}
\ds _0D_t^{\alpha}f = \frac{1}{\Gamma(1-\alpha)} \frac{\mathrm{d}}{\mathrm{d}t}\int_0^t\frac{f(t')}{(t-t')^{\alpha}}dt',\;\;\;0<\alpha<1,
\end{equation}	
where $\alpha$ is the fractional order and $\Gamma(\cdot)$ denotes the standard Gamma function.

\end{definition}

\begin{flushleft}
	\textbf{Note:} In this paper, if we take fractional order $\alpha=1$, the fractional derivative reduces to the classical first order derivative.
\end{flushleft}
The following semigroup property could be utilized to design a numerical algorithm later. Typically speaking for general cases, semigroup property is unvalid for Riemann-Liouville fractional operators. However, if the fractional orders are restricted in some special range, the property is valid.
\begin{property}
	\cite{Li} If $f(t)\in C^1[0,T],\;\alpha_i\in (0,1)\;(i=1,2)$, and $\alpha_1 +\alpha_2 \in (0,1]$, then $_0D_t^{\alpha_1}\cdot {_0D_t^{\alpha_2}f(t)} = {_0D_t^{\alpha_1+\alpha_2}f(t)}$.
\end{property}

\textbf{Notations:}
\begin{itemize}
	\item $I(t)$: cumulative infected people at time $t$;
	\item $J(t)$: cumulative confirmed people at time $t$;
	\item $G(t)$: currently isolated people who are infected but still in latent period at time $t$;
	\item $R(t)$: cumulative cured people at time $t$;
	\item $\beta$: spread rate;
	\item $\gamma$: morbidity;
	\item $\ell$:  the rate of isolation;
	\item $\kappa$: cure rate.
\end{itemize}

\textbf{Assumptions:}
\begin{itemize}
	\item [1.] The infected people averagely experience a latent period of $\tau_1$ days before they have obvious symptoms. It is assumed that once symptoms appearing, those infected people will seek treatment and therefore become diagnosed people.
	\item [2.] Because of the local government intervention, some of the infected people while still in the latent period would be isolated before they have significant symptoms. Assuming that they would be diagnosed in the next $\tau_1'$ days, which means that the average exposed period of these people are $\tau_1-\tau_1'$;
	\item [3.] We assume that a person once isolated or in treatment, the individual would no longer transmit the coronavirus to others;
	\item [4.] The confirmed people would take $\tau_2$ days in average to be cured with rate $\kappa$;
	\item [5.] We assume that growing process of $J(t)$ and $R(t)$ is a sub-diffusion process which means they depend on the corresponding history data from the start time to the moment.
\end{itemize}

Based on the above notations and assumptions, the dynamics of these popultaions $I(t),\;J(t),\;G(t),\;R(t)$ are described by the following fractional time delay dynamic (FTDD) system:
\begin{align}
&\label{1}\frac{\mathrm{d}I}{\mathrm{d}t}=\beta(I(t)-J(t)-G(t)),\\
&\label{2}_0D_t^{\alpha}J=\gamma\int_{0}^{t}h_{1}(t-\tau_{1},t')\beta(I(t')-J(t')-G(t'))\mathrm{d}t',\\
&\label{3}\frac{\mathrm{d}G}{\mathrm{d}t}=\ell(I(t)-J(t)-G(t))-\ell\int_{0}^{t}h_{2}(t-\tau_{1}',t')(I(t')-J(t')-G(t'))\mathrm{d}t',\\
&\label{4} _0D_t^{\lambda}R = \kappa \int_{0}^{t}h_{3}(t-\tau_{1}-\tau_2,t')\beta(I(t')-J(t')-G(t'))\mathrm{d}t'.
\end{align}


For easiness of understanding the system $\eqref{1}$-$\eqref{4}$, we present some explanations.

\begin{itemize}
	\item [1.] The cumulative confirmed people $J(t)$ all come from the infected people in the latent period $\tau_1$;
	\item [2.] The instant change in $G(t)$ has two factors. One is that some of infected people are isolated by the local government. Another factor is some isolated people are diagnosed after $\tau_1'$ days;
	\item [3.] $h_i(\hat{t},t')\;(i=1,2,3)$ is a distribution which should be normalized as
	\begin{equation*}
	\int_0^th_i(\hat{t},t'){\rm d}t' =1,\;\;t'\in (0,t),\;\;i=1,2,3.
	\end{equation*}
	Here we take the normal distribution
	\begin{equation*}
	h_i(\hat{t},t') = c_ie^{-c_{i+3}(\hat{t}-t')^2}
	\end{equation*}
	with $c_i$ and $c_{i+3}$ be constants.
\end{itemize}

\section{The Reconstruction and Prediction Algorithm}

In this section, a reconstruction algorithm of two parameters (spread rate and rate of isolation) of the dynamic system $\eqref{1}$-$\eqref{4}$ is introduced based on the optimization method, utilizing the official data.

According to Property 2.1, we take 1-$\alpha$ derivative on both side of the equation $\eqref{2}$ and 1-$\lambda$ derivative on both side of the equation $\eqref{4}$. Then we can transform the system $\eqref{1}$-$\eqref{4}$ into

\beq \left\{
\begin{array}{l}
	\ds \frac{\mathrm{d}I}{\mathrm{d}t}=\beta(I(t)-J(t)-G(t)),\\
	\vspace{.15cm} \ds \frac{\mathrm{d}J}{\mathrm{d}t}=\gamma\left({_0D_t^{1-\alpha}}( \int_{0}^{t}h_{1}(t-\tau_{1},t')\beta(I(t')-J(t')-G(t'))\mathrm{d}t' )\right), \\
	\vspace{.15cm} \ds \frac{\mathrm{d}G}{\mathrm{d}t}=\ell(I(t)-J(t)-G(t))-\ell\int_{0}^{t}h_{2}(t-\tau_{1}',t')(I(t')-J(t')-G(t'))\mathrm{d}t',\\
	\vspace{.15cm} \ds \frac{\mathrm{d}R}{\mathrm{d}t} = \kappa \left( _0D_t^{1-\lambda} ( \int_{0}^{t}h_{3}(t-\tau_{1}-\tau_2,t')\beta(I(t')-J(t')-G(t'))\mathrm{d}t' ) \right). \\
\end{array}\right.\label{31}\eeq
For given parameters $\{\beta,\gamma,\ell,\kappa,\tau_1,\tau_1',\tau_2,\alpha,\lambda\}$ and the initial conditons $\{I(t_0),G(t_0),J(t_0),R(t_0)\}$ of the above dynamic system $\eqref{31}$, we intend to apply the Matlab innner-embedded program \textbf{dde23} to solve the dynamic system numerically. Furthermore, the cumulative diagnosed people $J(T)$ and cumulative cured people $R(T)$ at any given time $T$ are easily to be obtained.

We can also adpot implicit Euler scheme to solov the dynamic system $\eqref{1}$-$\eqref{4}$ directly. Let $t_0<t_1<\cdots<t_N=T$, where $N$ is the total number of intervals. This is a uniform time step $\tau = T/N$. The shifted Gr$\rm\ddot{u}$nwald formula \cite{Liu} for approximationg the Riemann-Liouville derivative at time $t = t_n,\;(1\le k \le n)$ is given by
\begin{equation*}
_0D_t^{\alpha}f(t_n) \approx \frac{1}{\tau^{\alpha}}\sum_{j=0}^{n}w_{j}f_{n-j},
\end{equation*}
where
\begin{equation*}
w_0 =1,\;w_j = (-1)^{j}\frac{\alpha(\alpha-1)\cdots(\alpha-j+1)}{j!},\;\;j= 1,2,3,\ldots.
\end{equation*}
Let $\mu = \tau^{\alpha+1}\gamma\beta$. The following linear equations hold:
\beq \left\{
\begin{array}{l}
	\ds (1-\tau\beta)I_n + \tau\beta J_n + \tau\beta G_n = I_{n-1},\\
	\vspace{.15cm} \ds {\mu}h_1^nI_n + (w_0+{\mu}h_1^n)J_n + {\mu}h_1^n G_n = -w_nJ_0 + \sum_{j=1}^{n-1} \left({\mu}h_1^jI_j-(w_{n-j}+{\mu}h_1^j)J_j-{\mu}h_1^jG_j\right), \\
	\vspace{.15cm} \ds \tau\ell(\tau h_2^n-1)I_n + \tau\ell (1-\tau h_2^n)J_n + (1+\tau \ell -\tau^2\ell h_2^n) G_n = G_{n-1} - \tau^2 \ell \sum_{j=1}^{n-1}h_2^j(I_j-J_j-G_j),\\
	\vspace{.15cm} \ds \tau^{\lambda}\kappa\beta(-I_n+J_n+G_n) +  w_0R_n = \tau^{\lambda}\kappa\beta\sum_{j=1}^{n-1}(I_j-J_j-G_j) - \sum_{j=0}^{n-1}w_{n-j}R_j, \\
\end{array}\right.\label{32}\eeq
where $I_n,J_n,G_n,R_n$ are the numerical solutions of $I(t_n),J(t_n),G(t_n),R(t_n)$, and $h_i^n = h_i(t_n)$.

In addition, on the initial day $t_0$, it is assumed that 5 people were infected the COVID-19 from unknown sources, i.e., $I(t_0) = 5$. The confirmed, isolted and recoverer people is 0, i.e., $J(t_0) = G(t_0) = R(t_0) = 0$. Moreover, it is assumed that no isolation exists before $T=t+15$. According to the public data, we suppose a relatively high morbidity $\gamma = 0.99$ and cure rate as $\kappa = 0.97$. The average latent period $\tau_1$ and treatment $\tau_2$ are also regarded as known, i.e., $\tau_1 = 8$ and $\tau_2 = 12$. The average period between getting isolated and diagnosed $\tau_1'=5$.
Therefore, the parameters that need to be estimated are
\begin{equation*}
\theta:=[\beta,\ell]\;\;\;\rm and\;\;\;\Delta:=[\alpha,\lambda].
\end{equation*}
The identification of parameter $\theta^*$ comes to the following optimization problem:
\begin{equation}\label{33}
 \theta^* = {\min}_{\theta}\left\|J(\theta;t,\Delta)-J_{Obs}\right\|,
\end{equation}
where $J_{Obs}$ is the daily official data reported by the Chinese Center for Disease Control and Prevention. The pseudo code for solving this optimization problem is given in {\bf Algorithm 1}.
\begin{algorithm}[H]
	\caption{Reconstruction and Prediction with LM Method}
	\label{alg1}
	\begin{algorithmic}[1]
		\REQUIRE Initial conditons $\{I(t_0),G(t_0),J(t_0),R(t_0)\}$,\;parameters $\{\gamma,\kappa,\tau_1,\tau_1',\tau_2,\alpha,\lambda\}$,\;initial guess $\theta_0 = [\beta_0, \ell_0]$,\;observation data $J_{Obs}$,\;and maximum number iteration $M$;
		\ENSURE Predicted value $I(t),J(t),G(t),R(t)$ and optimization parameter $\theta^*$;
		
		\FOR{$1\le i \le M$}
		\STATE Compute $J(\theta_{i-1};t)$ by solving $\eqref{31}$ with Matlab inner-embedded program \textbf{dde23} or $\eqref{32}$ with Euler scheme;
		\STATE Caculate $\theta_i$ by solving equation $\eqref{33}$ with Levenberg-Marquad (LM) method;
		\ENDFOR
			
		\STATE Compute the predicted value $I(t),J(t),G(t),R(t)$ by repeating step 2 with $\theta^*$ obtained by step 3.
	\end{algorithmic}
\end{algorithm}


\section{Numerical Results}

In this section, we shall present some numerical experiments to verify the rationality and effectiveness of our proposed model. The data including the cumulative diagnosed people $J(t)$ and the cumulative cured people $R(t)$ is available from January 31 untill February 14, 2020. Moreover, the data is acquired from Chinese Center for Disease Control and Prevention.

\begin{example}

\rm It is significantly important to study the impact of different parameter $\Delta$ on the dynamic model $\eqref{31}$. Therefore, we shall provide different $\Delta$ to solve the optimization problem and analyze sentativity of the estimated parameters. The following three cases are considered.
\begin{itemize}
	\item [Case 1.] When $\alpha > \lambda$, Figure 1 shows the prediction results with different $\lambda$ and fixed $\alpha$. The reconstructed parameters are exhibited in Table 1.
	\item [Case 2.] If $\alpha = \lambda$, the trends of cumulative confirmed and cured people are depicted in Figure 2, and reconstructed parameters in Table 2.
	\item [Case 3.] Setting $\alpha < \lambda$, we get the tendency of the $J(t)$ and $R(t)$ in Figure 3 and recovered parameters in Table 3.
\end{itemize}

\end{example}

\begin{figure}[htp]
	\centering
	\subfigure[$\lambda$ = 0.9]{\includegraphics[width=5cm,height=3.5cm]{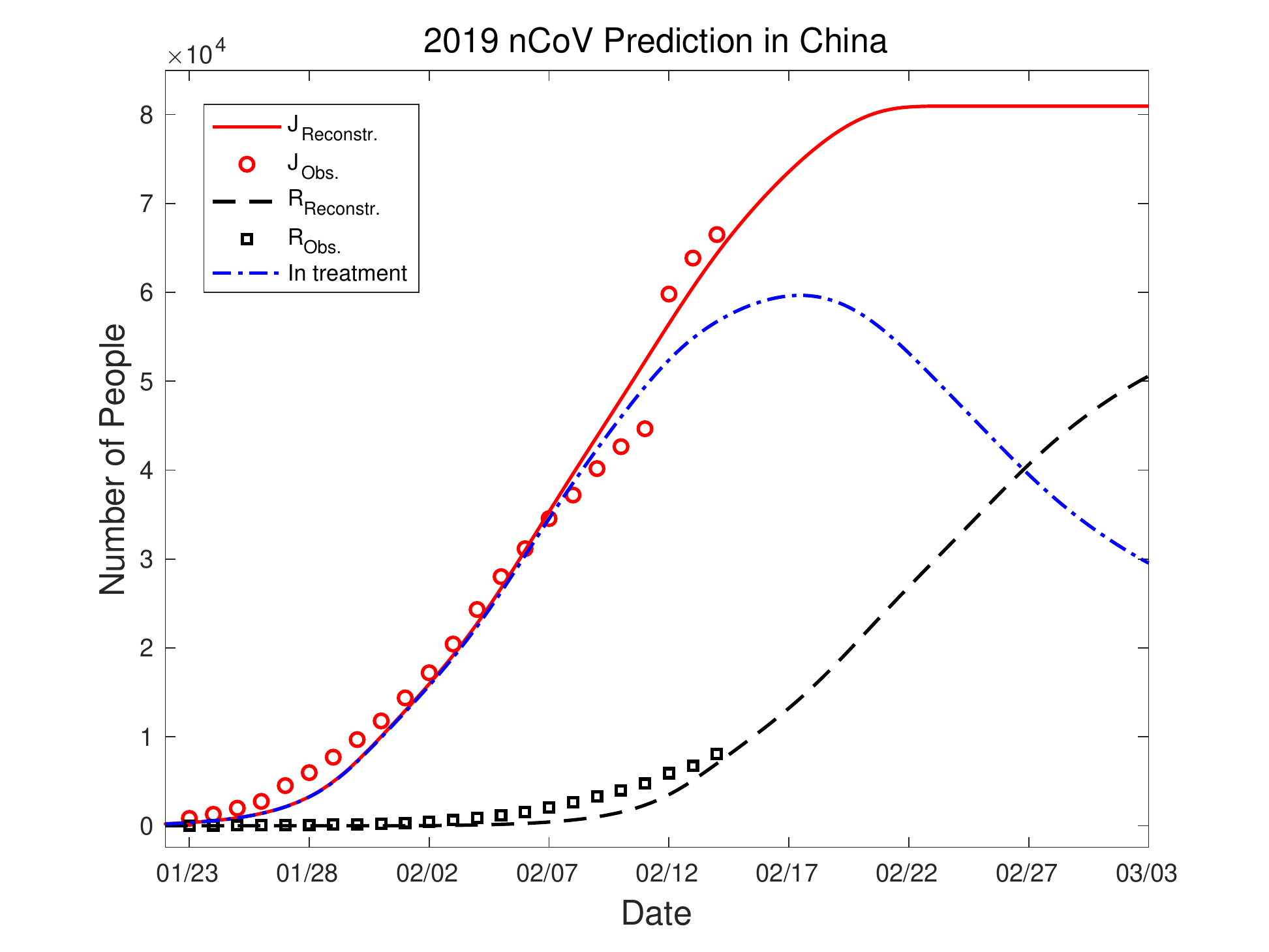}}
	\subfigure[$\lambda$ = 0.85]{\includegraphics[width=5cm,height=3.5cm]{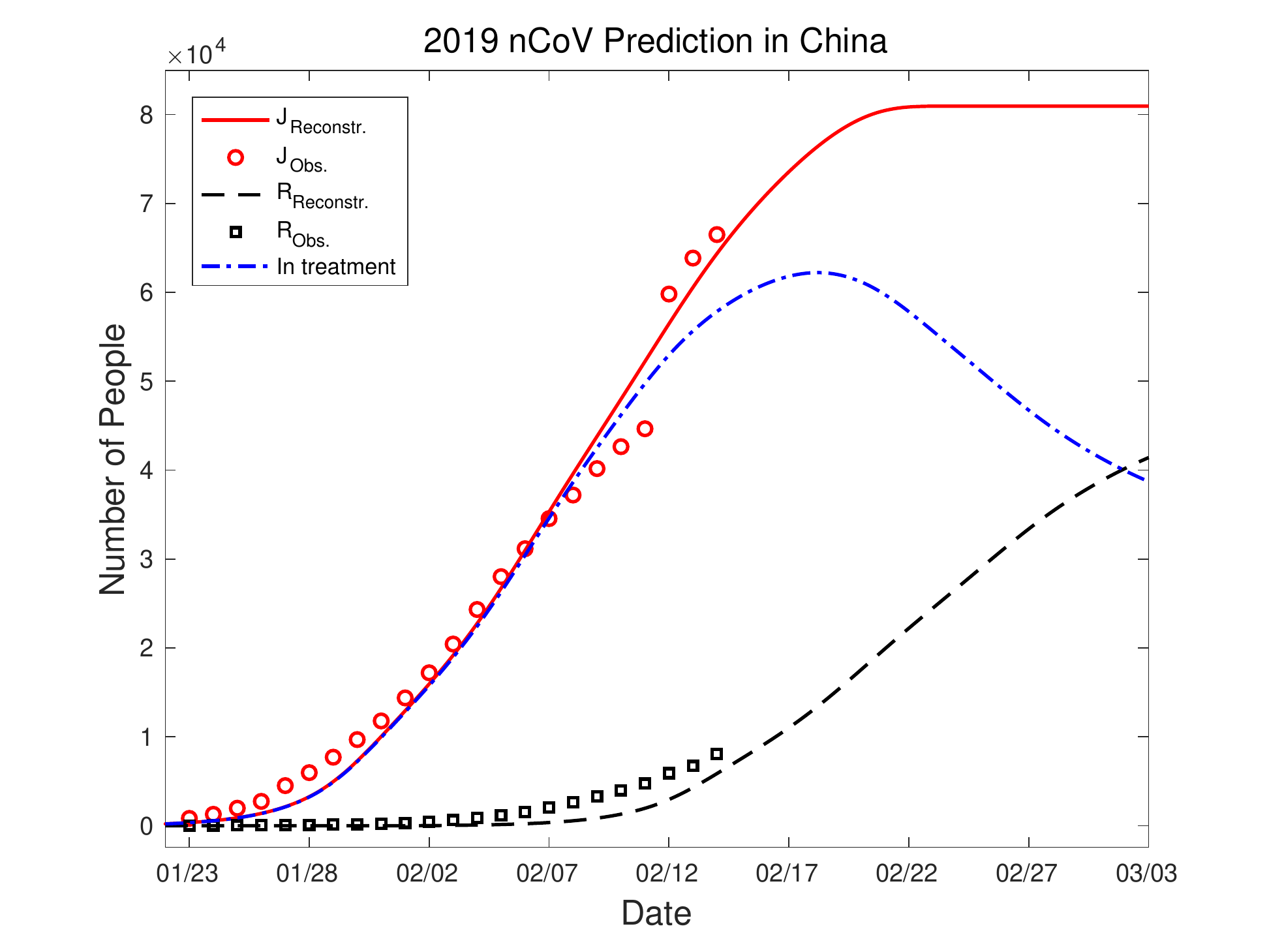}}
	\subfigure[$\lambda$ = 0.8]{\includegraphics[width=5cm,height=3.5cm]{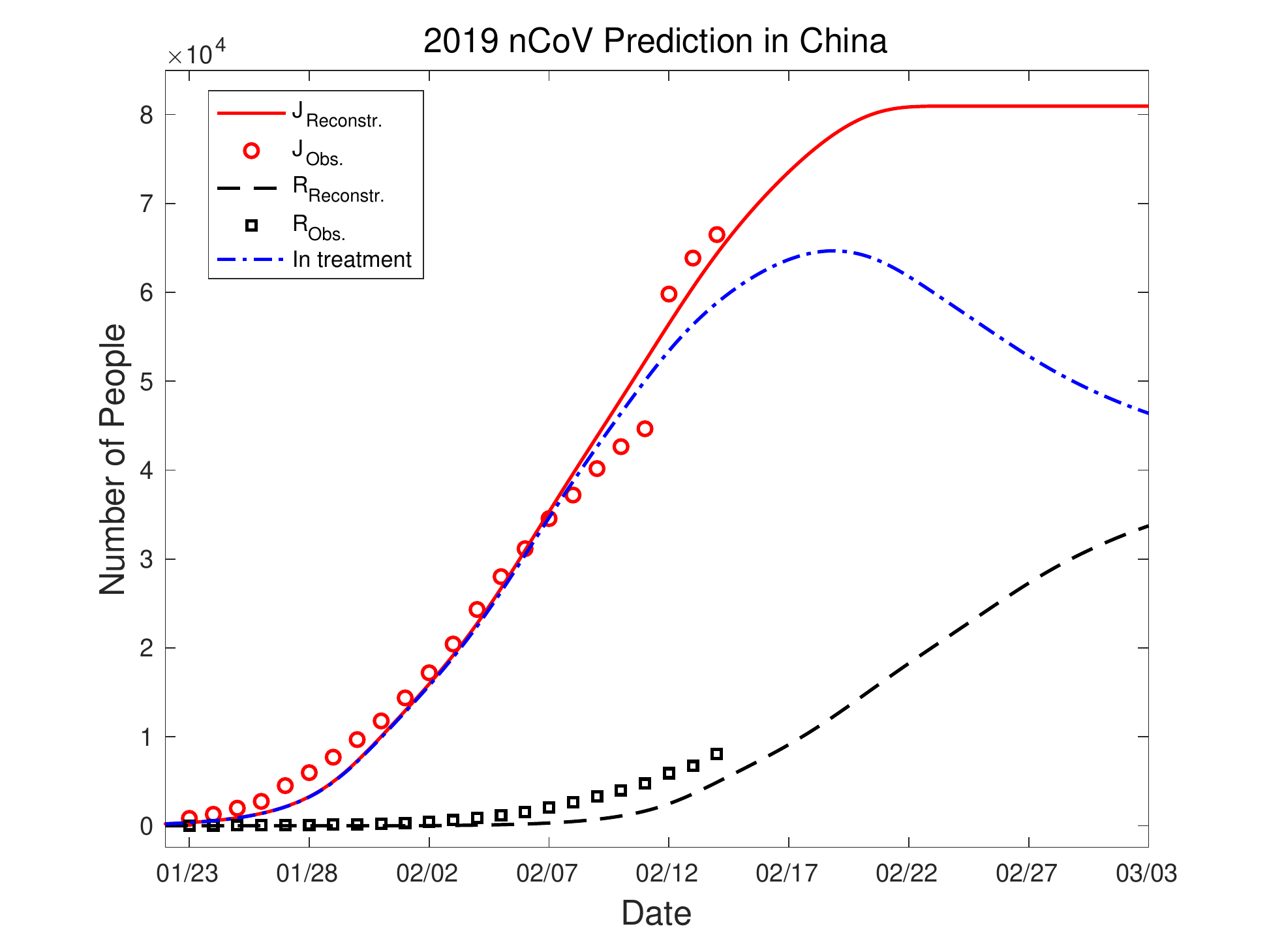}}
	\caption{Prediction with different $\lambda$ when $\alpha = 1$.}
	\label{fig1}
\end{figure}

\begin{table}[H]
	\begin{center}
		\caption{Estimated parameters}
		\begin{tabular}{cccc}
			\toprule
			$\alpha$ & $\lambda$  & $\beta$ & $\ell$ \\ \hline
			1.00 & 0.90 & 0.4222 & 0.4605  \\
			1.00 & 0.85 & 0.4222 & 0.4605 \\
			1.00 & 0.80 & 0.4222 & 0.4605 \\
			\toprule
		\end{tabular}
	\end{center}
	
	\label{tab1}
\end{table}

\begin{figure}[H]
	\centering
	\subfigure[$\alpha = 1,\;\lambda$ = 1]{\includegraphics[width=5cm,height=3.5cm]{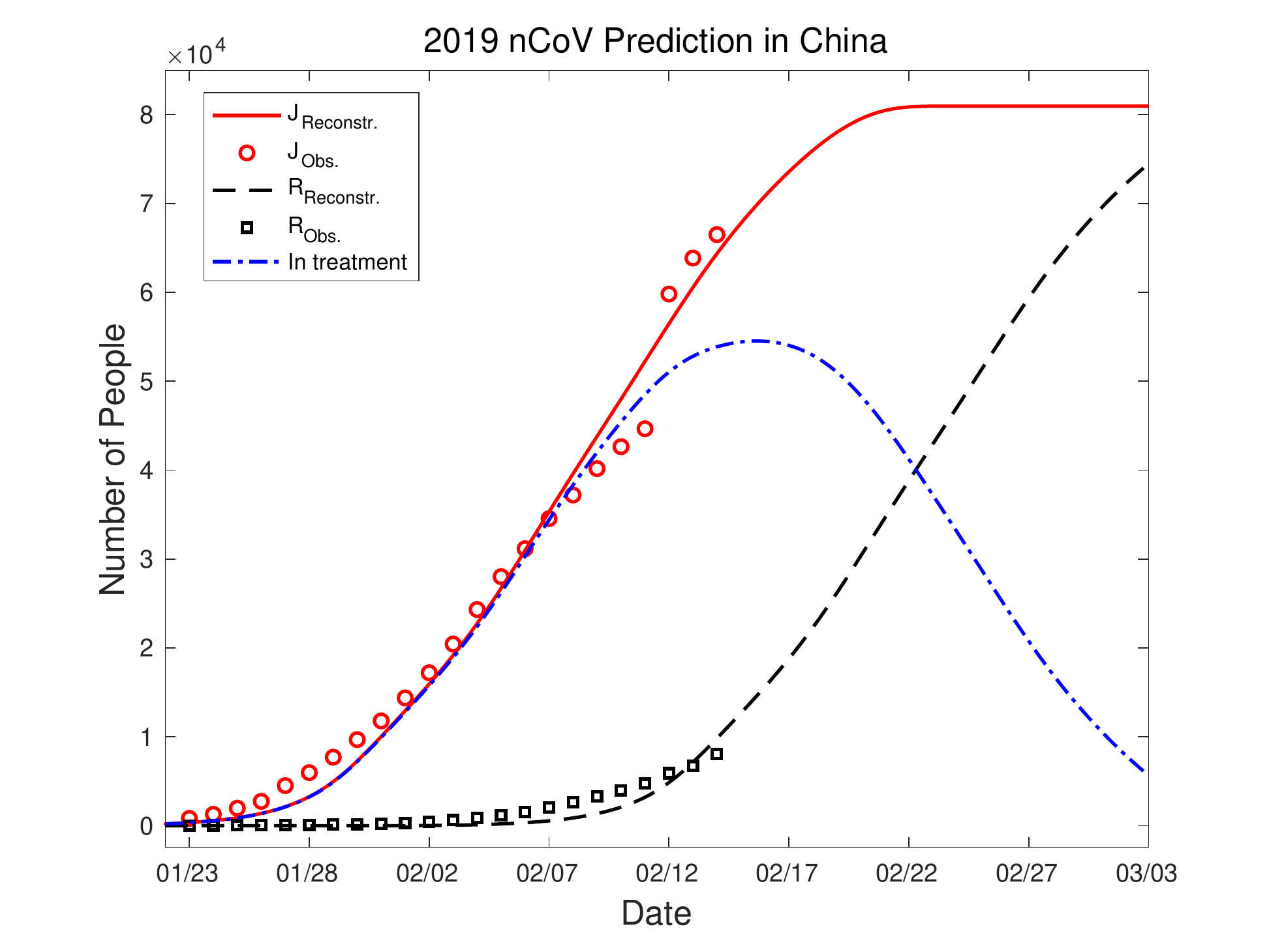}}
	\subfigure[$\alpha = 0.98,\;\lambda$ = 0.98]{\includegraphics[width=5cm,height=3.5cm]{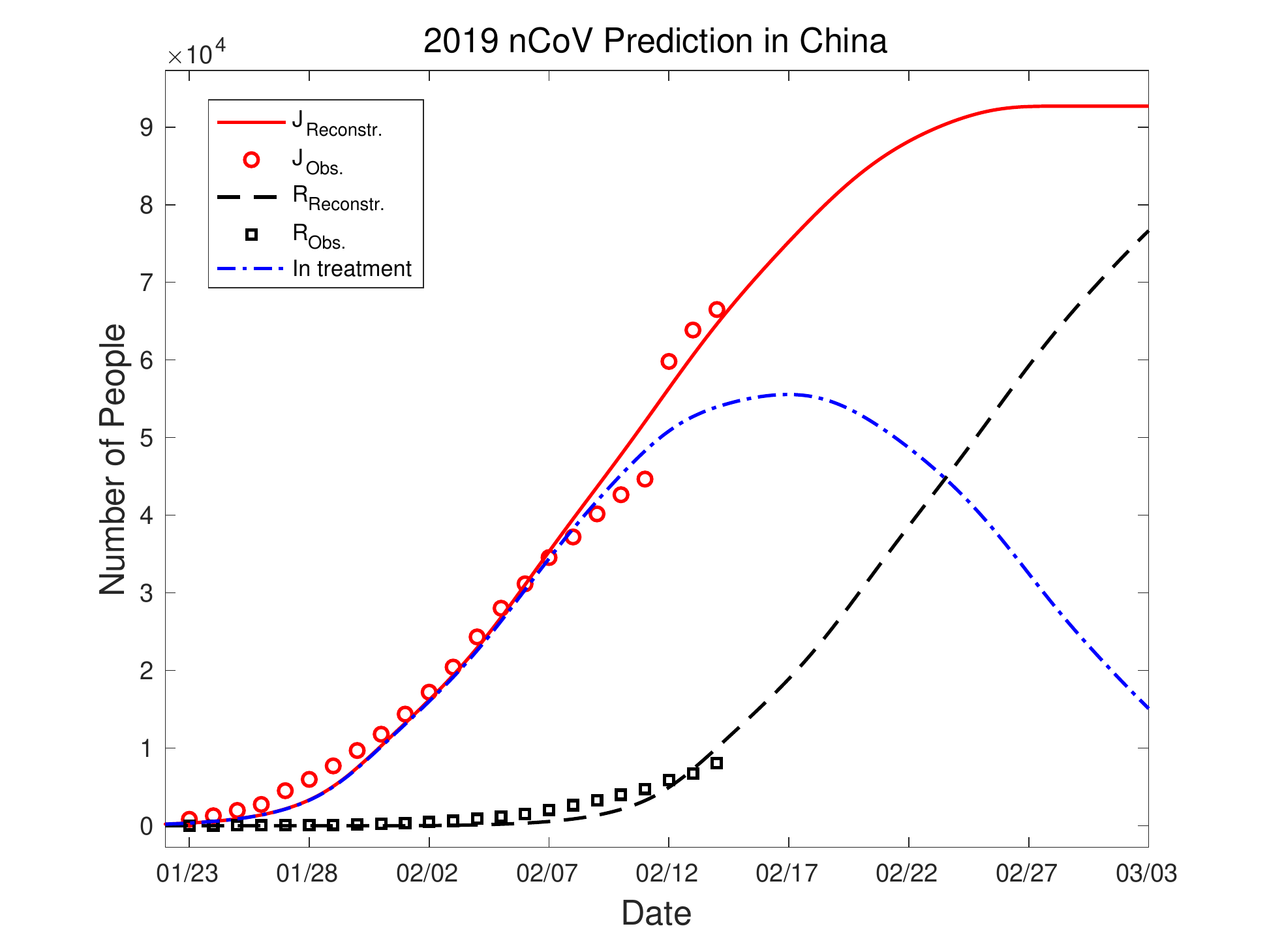}}	
	\subfigure[$\alpha = 0.96,\lambda$ = 0.96]{\includegraphics[width=5cm,height=3.5cm]{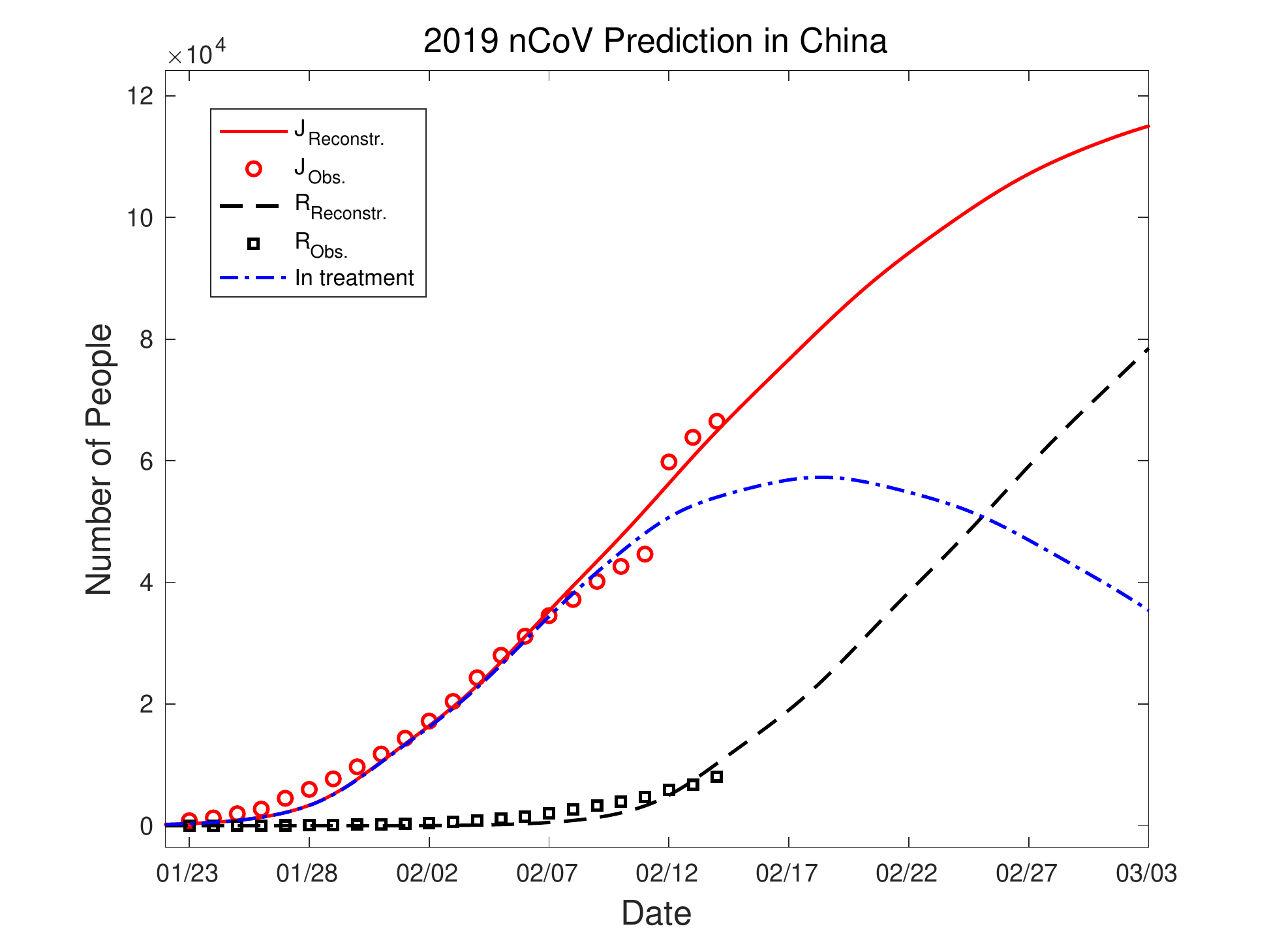}}
	\caption{Prediction with different $\alpha$ and $\lambda$ while $\alpha = \lambda$.}
	\label{fig2}
\end{figure}

\begin{table}[H]
	\begin{center}
		\caption{Estimated parameters}
		\begin{tabular}{cccc}
			\toprule
			$\alpha$ & $\lambda$  & $\beta$ & $\ell$ \\ \hline
			1.00& 1.00 & 0.4222 & 0.4605  \\
			0.98& 0.98 & 0.4296 & 0.4936 \\
			0.96& 0.96 & 0.4369 & 0.5265 \\
			\toprule
		\end{tabular}
	\end{center}
	
	\label{tab1}
\end{table}

\begin{figure}[H]
	\centering
	\subfigure[$\alpha = 0.95$]{\includegraphics[width=5cm,height=3.5cm]{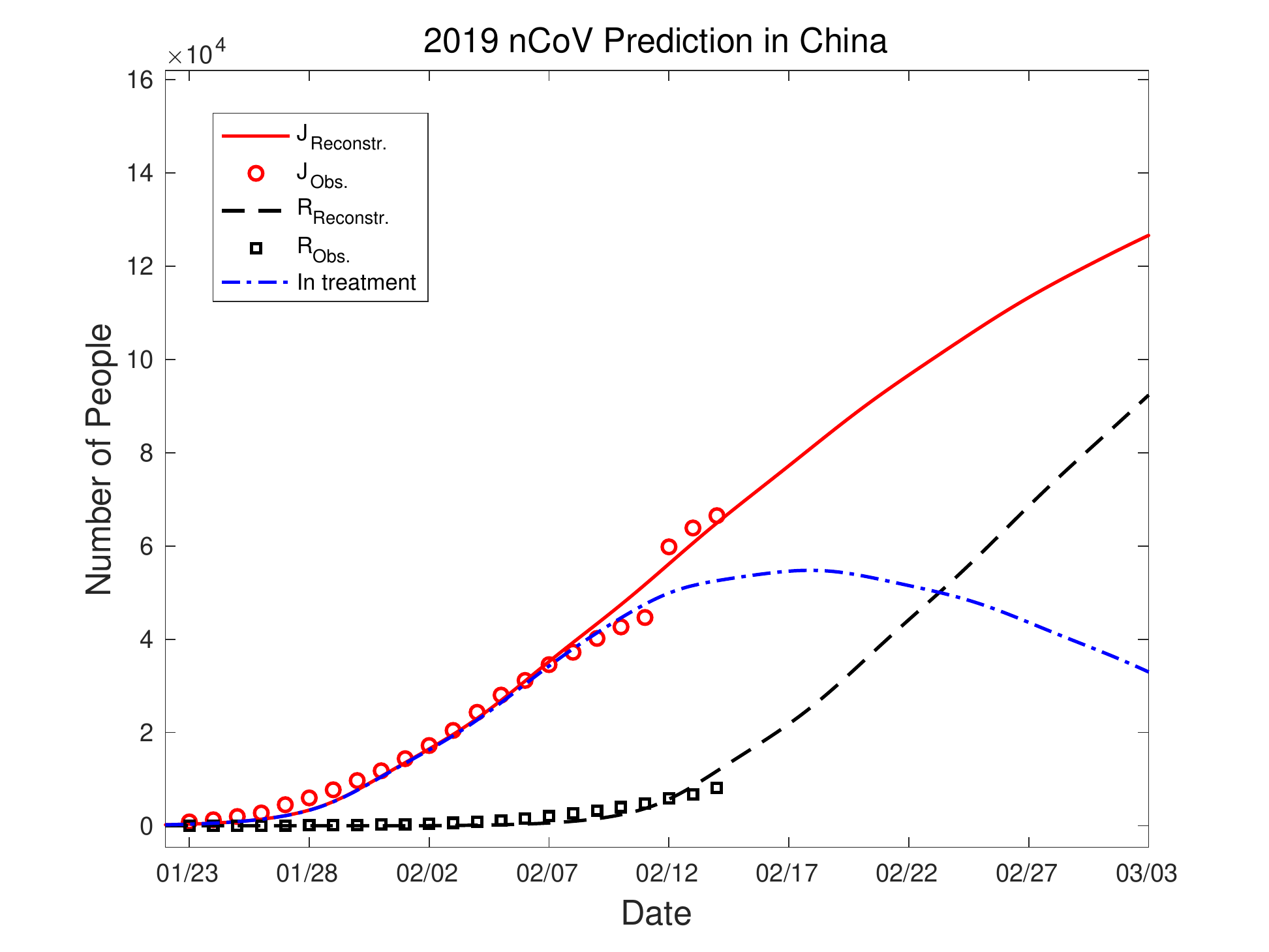}}
	\subfigure[$\alpha = 0.965$]{\includegraphics[width=5cm,height=3.5cm]{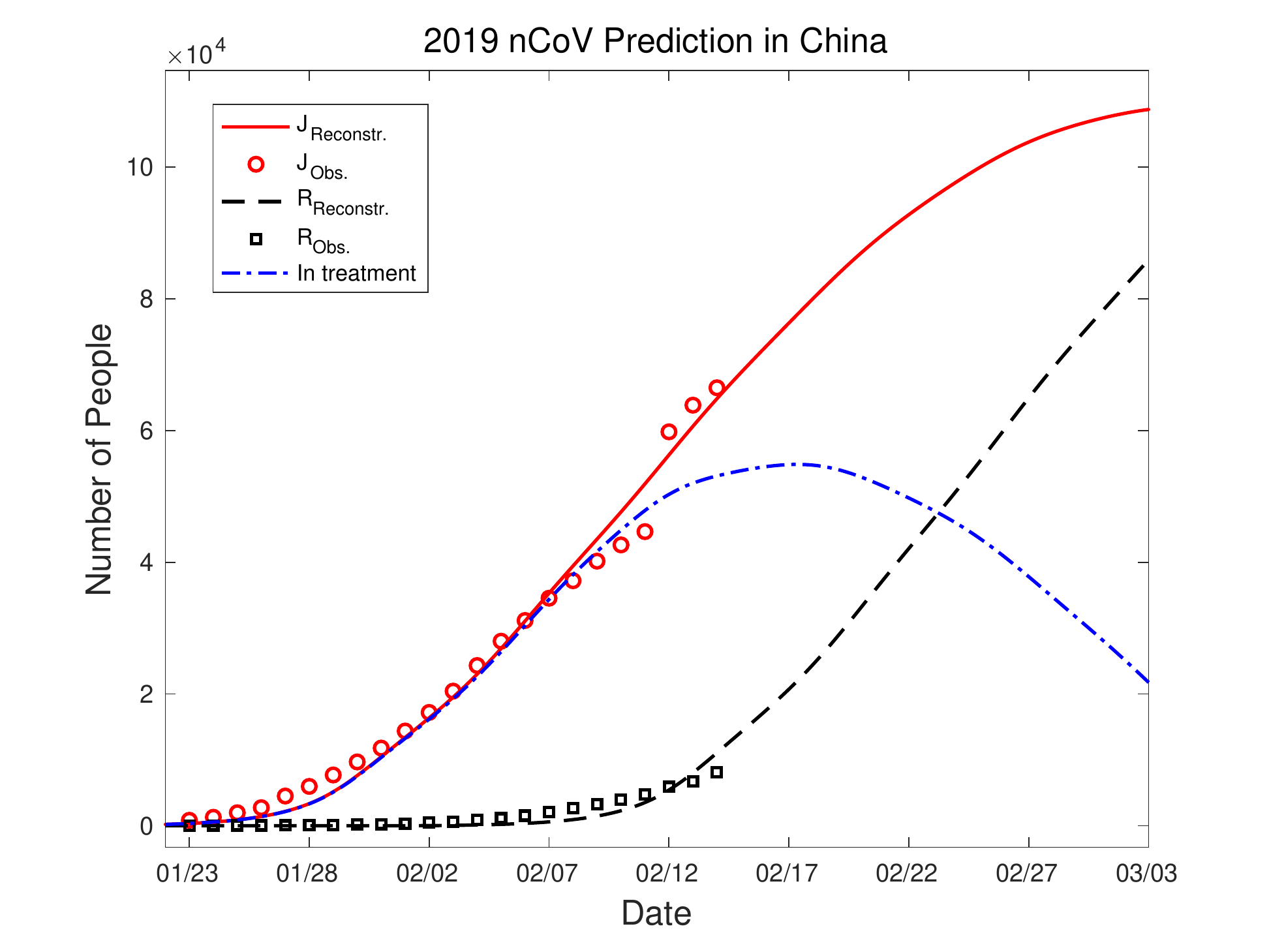}}
	\subfigure[$\alpha = 0.98$]{\includegraphics[width=5cm,height=3.5cm]{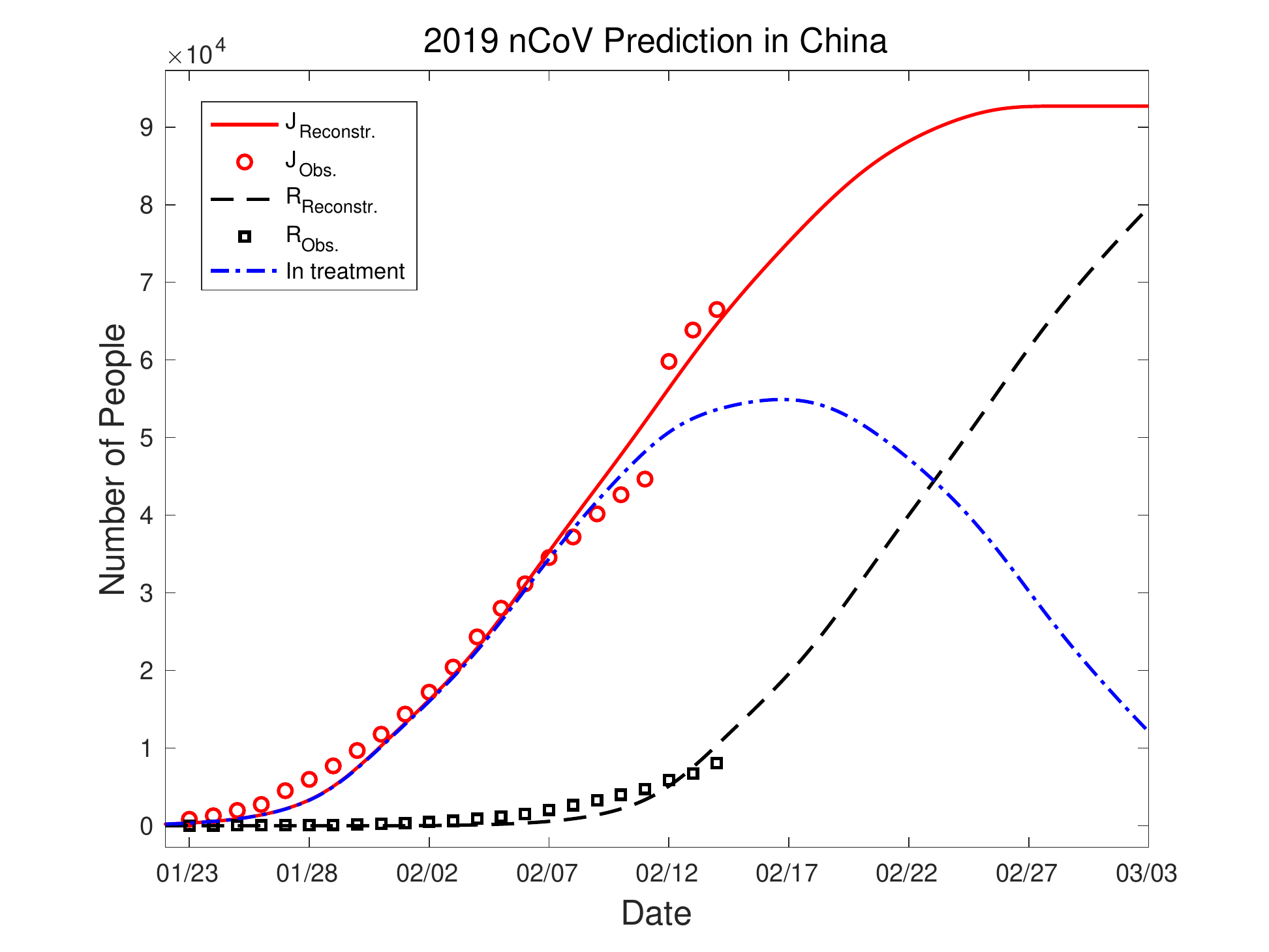}}	
	\caption{Prediction with different $\alpha$ when $\lambda =0.99$}
	\label{fig3}
\end{figure}

\begin{table}[H]
	\begin{center}
		\caption{Estimated parameters}
		\begin{tabular}{cccc}
			\toprule
			$\alpha$ & $\lambda$  & $\beta$ & $\ell$ \\ \hline
			0.95& 0.99 & 0.4351 & 0.5183 \\
			0.965& 0.99  & 0.4404 & 0.5428 \\
			0.98& 0.99 & 0.4296 & 0.4936 \\
			\toprule
		\end{tabular}
	\end{center}
	
	\label{tab3}
\end{table}

$J_{Obs.}$ and $R_{Obs.}$ are the data of cumulative diagnosed and cured people for reconstructions. $J_{Rconstr.}$ and $R_{Reconstr.}$ correspond to the estimated values of diagnosed and cured cases. Figure 1-3 and Table 1-3 show $\alpha$ is quite sensitive as the cumulative confirmed people are sharply increasing with slight decrease of $\alpha$.
The estimated number of confirmed people still increase and cannot be in a stable state on early March when $\alpha<0.965$. It is worthy to mention that there is no change on the value of $\beta$ and $\ell$ with the reduction of $\lambda$ shown in Table 1. However, the estimated number of cured people is reducing to the value that even less the people who are in treatment when $\lambda <0.85$.

\begin{remark}
	
	Figure 2(a) describes the predicted trend with integer order dynamic model. Comparing this results with other fractional order dynamic model, e.g., Figure 1(a), Figure 2(b) and Figure 3(b), we are easy to find that the estimated cumulative confirmed people $J_{Reconstr.}$ with fractional order are more than that with interger order. On the other hand, the estimated cuumulative cured people $R_{Reconstr.}$ with fractional order are less than that with interger order.
	
	In fact, many infected people are not easy to be diagnosed by Nucleic acid detection, which leads to the number of confirmed people less than real number of people who should be diagnosed. People who are cured may still relapse. The predicted results of fractional order dynamic model can exactly explain this phenomenon. In addition, we infer that $0.965<\alpha\le 1$ and $0.84 <\lambda\le 1$ are more appropriate.
	
\end{remark}

\begin{example}
	\rm Let $\alpha = 0.97,\;\lambda=0.87$. We provide some numerical results and reconstructed parameters with official data of two areas. One area is the Hubei Province. Another area is the outside of Hubei Provice while still in China. Besides, in order to verify the accuracy of the chosen fractional orders, we use the latest official data (from February 15 to February 19) to compare the results of prediction. The predicted values of confirmed and cured people are shown in Figure 4. In this figure, $J_{Reconstr.}$ and $R_{Reconstr.}$ are the estimated values of cumulative diagnosed and cured people for prediction. $J_{obs}$ and $R_{obs}$ correspond to the data of cumulative confirmed and cured people for reconstructions. $J_{test}$ and $R_{test}$ are displayed for comparison with the prediction.
\end{example}

\begin{figure}[H]
	\centering
	\subfigure[Outside of Hubei]{\includegraphics[width=6.5cm,height=4cm]{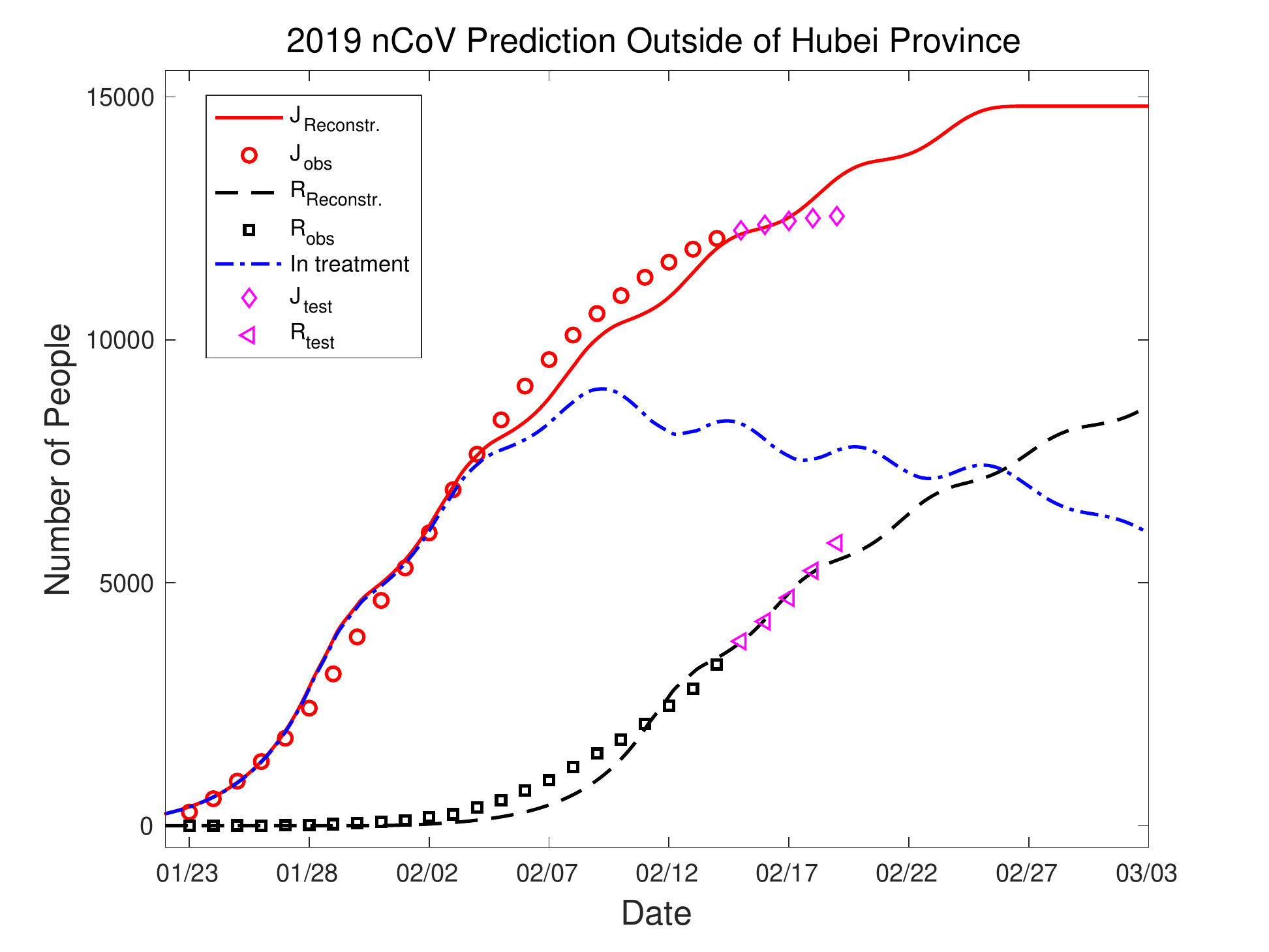}}
	\subfigure[Hubei Province]{\includegraphics[width=6.5cm,height=4cm]{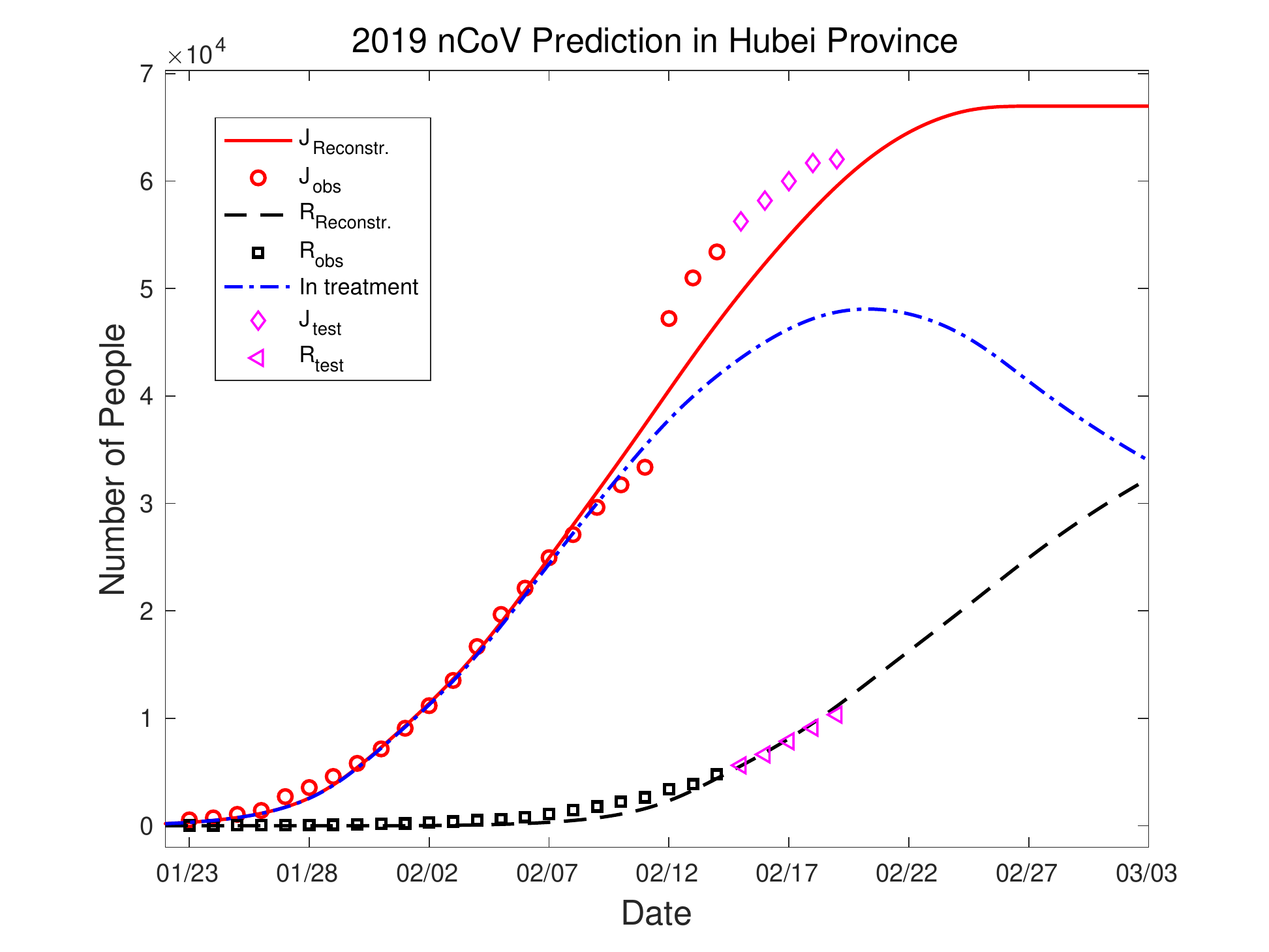}}
	\caption{Prediction and Comparison}
	\label{fig3}
\end{figure}

\section{Conclusions and Acknowledgement}

In this paper, we propose a novel time delay dynamic system with fractional derivative, where the growth of cumulative confirmed and cured people at time $t$ are both considered as a sub-diffusion process. Numerical simulations are carried out to verify the rationality and effectivness of the dynamic system.In addition, two proper interval are given for the fractional orders $\alpha$ and $\lambda$, respectively. Numerical examples show that the proposed fractional time delay dynamic system can generate prediction which agrees the public data well.

We are sincerely grateful to every member in our group for invaluable discussions, special to Prof. Yu Jiang, Prof. Jike Liu from Shanghai University of Finance and Economics and Prof. Wenbin Chen from Fudan University.  Jin Cheng was supported in part by the NSFC 11971121. Xiang Xu was partially supported by NSFC 11621101, 91630309 and the Fundamental Research Funds for the Central Universities.

\appendix
\makeatletter
\def\@seccntformat#1{\csname Pref@#1\endcsname \csname the#1\endcsname\quad}
\def\Pref@section{Appendix~}
\makeatother

\end{document}